\documentclass[cits]{PoS}
\pdfoutput=1
\title{Chiral restoration of the momentum space quark propagator through Dirac low-mode truncation}

\ShortTitle{Chiral restoration of the mom. space quark prop. through Dirac low-mode truncation}

\author{\speaker{Mario Schr\"ock}\\
        Institut f\"ur Physik, FB Theoretische Physik, Universit\"at
Graz, A--8010 Graz, Austria\\
       E-mail: \email{mario.schroeck@uni-graz.at}}

\abstract{
We calculate the chirally improved (CI) Landau gauge quark propagator in two flavor lattice Quantum Chromodynamics and study thereon the effects of Dirac low-mode removal. The application of tree-level improvement of the propagator and tree-level correction of the lattice dressing functions removes the leading discretization artifacts. We find the dynamically generated mass in the infrared domain of the mass function to dissolve continuously with the reduction level and furthermore we observe strong suppression of the wave-function renormalization function for small momenta.
}

\FullConference{Xth Quark Confinement and the Hadron Spectrum\\
		 8--12 October 2012\\
		 TUM Campus Garching, Munich, Germany}

\usepackage[tight]{units}
\usepackage{amsmath}
\usepackage{wrapfig}

\newcommand{\eq}[1]{(\ref{#1})}
\newcommand{\fig}[1]{Fig.~{\ref{#1}}}

\newcommand{\RD}[1]{{\mathrm{red}(#1)}}

\newcommand{\LM}[1]{{\mathrm{lm}(#1)}}

\newcommand{\ket}[1]{\left|{#1}\right>}
\newcommand{\bra}[1]{\left<{#1}\right|}
\newcommand{\be}{\begin{equation}}
\newcommand{\ee}{\end{equation}}
\newcommand{\gaf}{\gamma_5}

\newcommand{\bigO}{\mathcal{O}}

\newcommand{\kslash}{k\hspace{-2mm}\slash}
\newcommand{\pslash}{p\hspace{-0.475em}\slash}
\newcommand{\DCI}{D_{\textrm{CI }}}

\begin{document}

\section{Motivation}

\begin{figure}[htb]
	\center
	\includegraphics[width=0.85\textwidth]{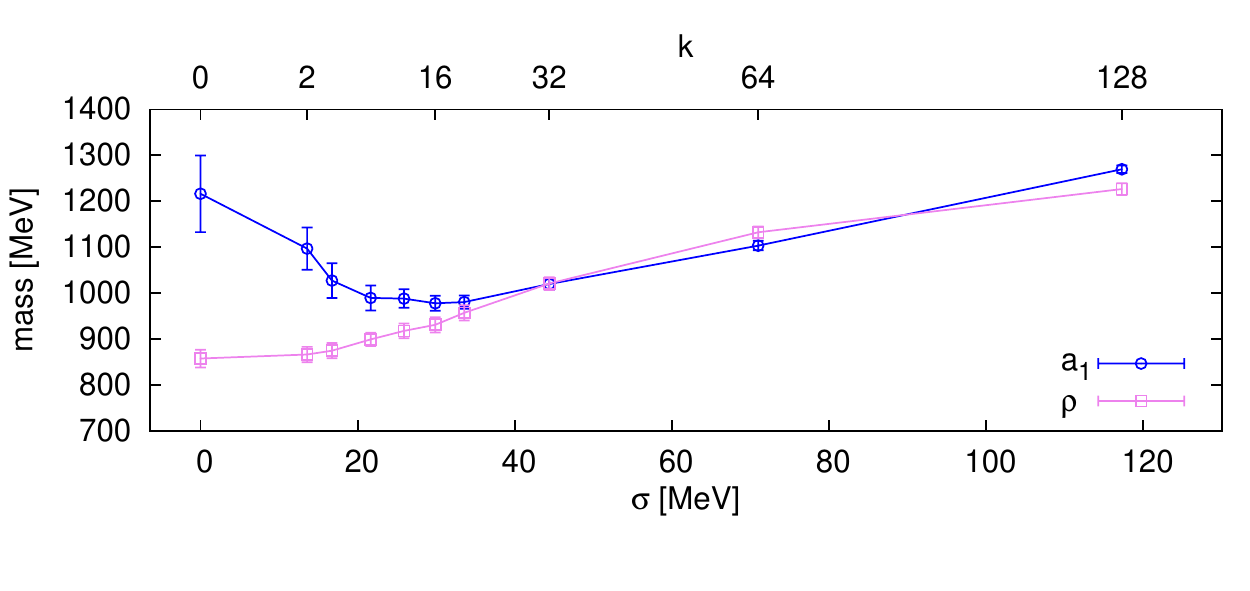}
	\caption{The masses of the vector meson $\rho$ and axial vector meson $a_1$
		under Dirac low-mode removal: the would be chiral partners
		become degenerate once the chiral condensate has been removed.}
	\label{fig:mesons}
\end{figure}

The lowest eigenmodes of the Dirac operator are related to the dynamical
breaking of the chiral symmetry in QCD via the Banks--Casher relation.
In \cite{Lang:2011vw, Glozman:2012fj} we investigated the effects of 
the removal of the lowest Dirac eigenmodes from the valence quark sector 
on the hadron mass spectrum in a dynamical lattice QCD simulation.
It was found that all hadrons, except for the pions, survive the aforementioned  
truncation. The masses of hadrons that can be transformed into each other by
a chiral rotation became degenerate and surprisingly the value of these masses remained 
rather large: see for example \fig{fig:mesons} where we show the evolution
of the vector and axial vector meson masses under Dirac low-mode removal.

The removal of the lowest Dirac eigenmodes, i.e., the removal of the chiral 
condensate is expected to result in the loss of dynamically generated mass
in the valence quarks. Thus, the naive expectation suggests rather light hadrons
once the chiral symmetry has been restored.
The study \cite{Schrock:2011hq} aimed at shedding some light on the issue of
mass generation in the context of Dirac low-mode removal: here we explored
the valence quark propagators that have been used in the previous studies,
in a gauge fixed setting in momentum space. Then it is possible to extract
at each Dirac low-mode truncation level
the mass function of the propagator which manifestly
shows the amount of dynamically generated mass in the quarks.

\section{Dirac low-mode truncation}
The lowest eigenmode of the Dirac operator are known to play a crucial role 
in the dynamical breaking of the chiral symmetry as stated by the Banks--Casher 
relation.
Consequently, when removing the Dirac eigenmodes near the origin,
the chiral condensate disappears and chiral symmetry becomes 
``unbroken'' \cite{Lang:2011vw}.

Consider the Hermitian Dirac operator $D_5\equiv \gamma_5D$.
We denote its (real) eigenvalues with $\mu_i$ and the corresponding eigenvectors with $\ket{v_i}$.
We write the Dirac operator $D$ in terms of the spectral representation of $D_5$:
\be\label{gafDspectral}
	D = \sum_{i=1}^N \,\mu_i\,\gamma_5\,\ket{v_i}\bra{v_i}.
\ee
Then we can split the quark propagator $S=D^{-1}$ into a low-mode part (lm) 
and a reduced part (red),
\be
	S = \sum_{i\leq k}\, \mu_i^{-1} \,\ket{v_i}\bra{v_i}\,\gaf 
	  + \sum_{i>k}\,\mu_i^{-1}\,\ket{v_i}\bra{v_i}\,\gaf
	  \equiv S_\LM{k} + S_\RD{k}
\ee
and hence we obtain the reduced part of the propagator by simply subtracting 
the low-mode part from the full propagator
\be\label{SRD}
	S_\RD{k} = S - S_\LM{k}.
\ee
It is this reduced part of the propagator which we calculate within a gauge fixed
setting in order to extract its momentum space dressing functions.

We use dynamical two flavor gauge field configurations \cite{Gattringer:2008vj, Engel:2010my}
for our study where for the dynamical quarks as well as for the valence quark propagator that we investigate,
the so called chirally improved (CI) Dirac operator \cite{Gattringer:2000js, Gattringer:2000qu} 
has been adopted.
We gauge fixed the configurations to Landau gauge with the \emph{cuLGT} code \cite{Schrock:2012fj} 
for lattice gauge fixing on graphic processing units (GPUs) with CUDA.

\section{The CI quark propagator from the lattice}
The continuum quark propagator at tree-level reads
\be\label{treelevelcont}
	S^{(0)}(p) = \left(i\pslash +m\right)^{-1}
\ee
where $m$ denotes the bare quark mass.
In manifestly covariant gauges the interacting 
renormalized quark propagator $S(\mu;p)$
can be decomposed into a Dirac scalar and a Dirac vector part
\be
	S(\mu;p) = \left(i\pslash A(\mu;p^2)+B(\mu;p^2)\right)^{-1}
\ee
or equivalently
\be\label{fullcont}
	S(\mu;p) = Z(\mu;p^2)\left(i\pslash +M(p^2)\right)^{-1}.
\ee
Here we introduced
the wave-function renormalization function $Z(\mu;p^2)=1/A(\mu;p^2)$ and
the mass function $M(p^2)=B(\mu;p^2)/A(\mu;p^2)$.

The regularized lattice quark propagator
$S_L(p;a)$ can be renormalized at the renormalization point $\mu$ 
with the quark wave-function renormalization constant $Z_2(\mu;a)$,
\be
	S_L(p;a) = Z_2(\mu;a) S(\mu;p).
\ee

The mass function $M(p^2)$ is independent of the renormalization point $\mu$ 
whereas the wave-function renormalization function $Z(\mu;p^2)$
differs at different scales although can be related from different scales
by multiplication with a constant.

We extract the functions $M(p^2)$ and $ Z(p^2)$ of the CI quark propagator
from a the lattice. For details of the extraction see Ref.~\cite{Skullerud:1999gv}.

The lattice quark propagator at tree-level
\be\label{treelevellat}
	S_L^{(0)}(p) = \left(ia\kslash +aM_L^{(0)}(p)\right)^{-1}
\ee
differs from the continuum case due to discretization artifacts.
The dressing function $A_L^{(0)}(p)$ equals one,  by construction,
and hence $B_L^{(0)}(p)$ equals 
at tree-level the mass function $M_L^{(0)}(p)$.

\begin{figure}[htb]
	\center
	\includegraphics[width=0.42\textwidth]{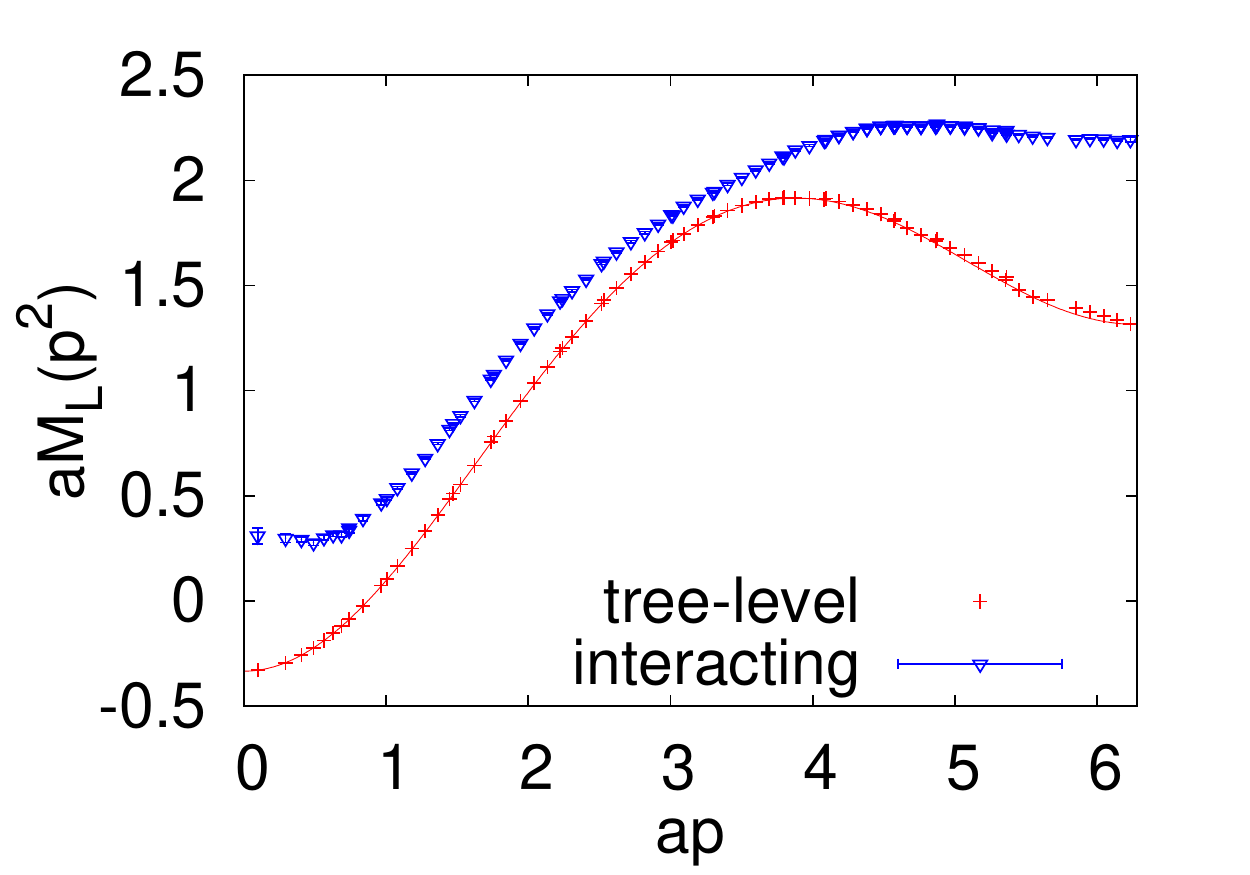}
	\includegraphics[width=0.42\textwidth]{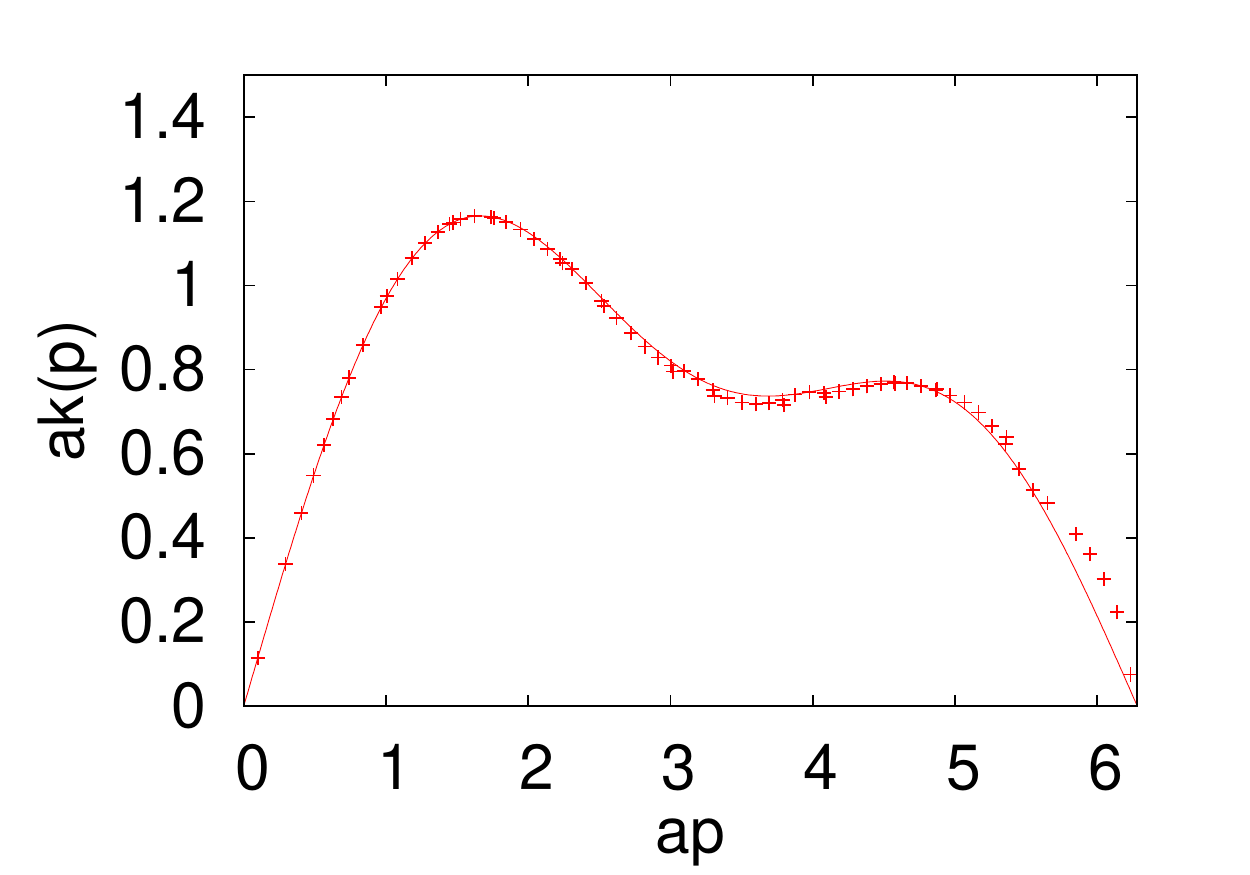}
	\caption{The tree-level mass function (left in red) and
		lattice CI momenta (right) together with the corresponding analytical
		curves and the uncorrected interacting mass function (left in blue).}
	\label{fig:treelevel}
\end{figure}

We extract the CI lattice momenta $ak(p)$ from the tree-level propagator and
compare it to the analytically derived expression for
the $\DCI$ momenta given in \cite{Schrock:2011hq}.
The lattice calculation is consistent with the analytical result, see
the r.~h.~s. plot of \fig{fig:treelevel}.

The tree-level mass function $aM_L^{(0)}(p)$ which in the continuum  is simply the bare mass $m$,
is shown in the l.~h.~s. plot of the same \fig{fig:treelevel} (in red), again consistency with the 
corresponding analytical expression is manifest.
As can be seen from the plot, $aM_L^{(0)}(p)$ has a zero-crossing with $aM_L^{(0)}(0)\approx -0.333$.

When switching from the tree-level case to the non-perturbative full interacting case,
the mass function $aM_L^{(0)}(p)$ of the l.~h.~s. plot of \fig{fig:treelevel} (red)
turns into the interacting lattice mass function $aM_L(p)$ (blue). 

The shape of $aM_L(p)$ is similar to $aM_L^{(0)}(p)$ and 
is clearly altered by discretization errors.
To get a handle on the leading lattice artifacts we perform
improvement and tree-level correction.
For the application of the Symanzik improvement program \cite{Symanzik:1983dc}
to reduce the errors of the fermionic action to $\bigO(a^2)$
we refer to \cite{Schrock:2011hq} and references therein.

In order to blank out the lattice artifacts which are already present
at tree-level, we aim at extracting the deviation of the interacting propagator
from its tree-level form. To this end we simply divide
the renormalization function $Z_L(p)$  by
its tree-level form
\be
	Z_L(p) \to\frac{Z_L(p)}{Z_L^{(0)}(p)}.
\ee
In order to apply an equivalent tree-level correction 
to the mass function of the form
\be
	aM_L(p)\to \frac{amM_L(p)}{M_L^{(0)}(p)}
\ee
we have to carry out an additive mass renormalization of the tree-level function $B_L^{(0)}(p)$ 
in order to avoid divergences because of the zero crossing of the tree-level function,
\be
	aB_L^{(0)}(p)\to aB_L^{(0)}(p)+ am_\textrm{add}
\ee
where $am_\textrm{add}$ is chosen such that $B_L^{(0)}(0)=m$, like in the continuum, i.e.
\be
	am_\textrm{add}=am-aB_L^{(0)}(0) \approx 0.344.
\ee
Thus, the multiplicative tree-level correction 
for the mass function is
\be 
	aM_L(p)\to\frac{amM_L(p)A_L^{(0)}(p)}{B_L^{(0)}(p)+m_\textrm{add}}.
\ee

\section{Results}
\begin{figure}[htb]
	\center
	\includegraphics[width=0.49\textwidth]{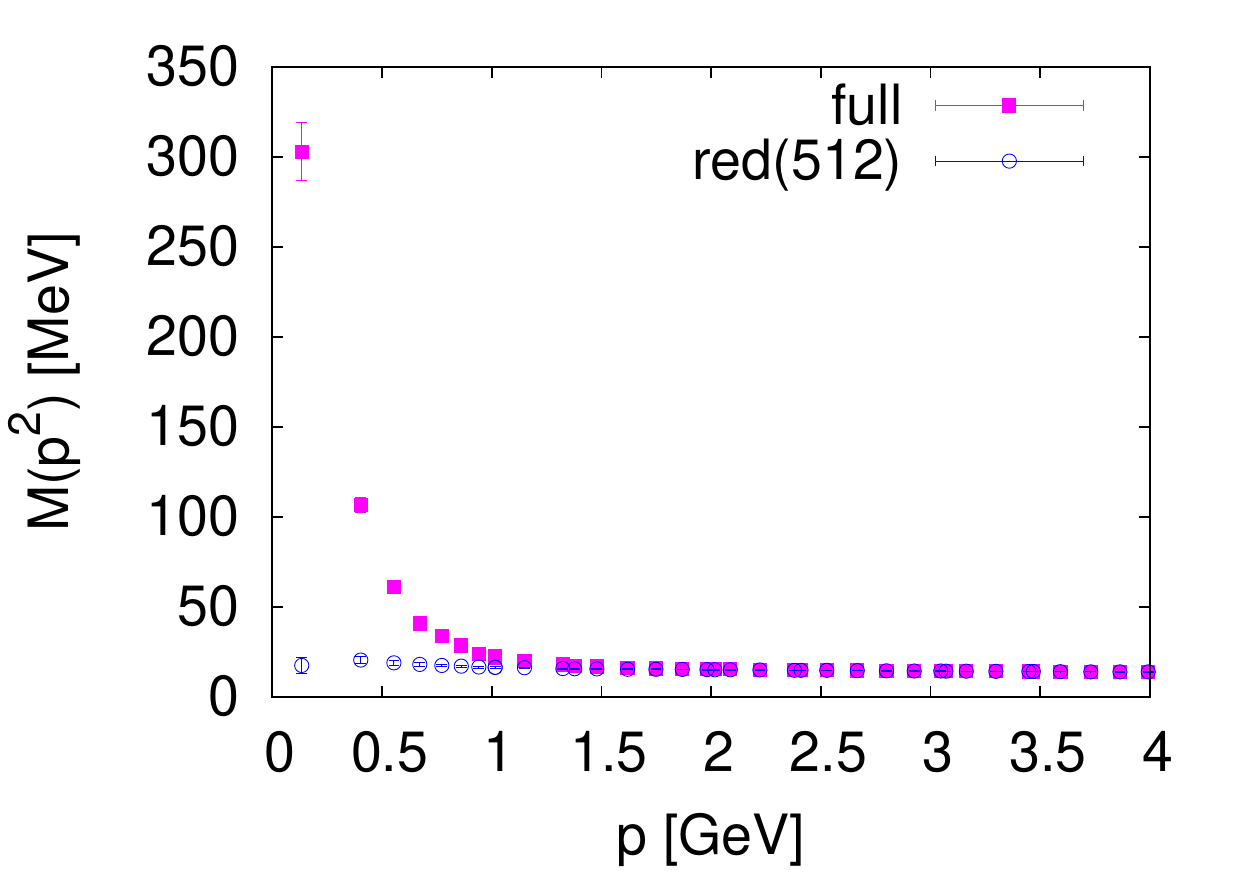}
	\includegraphics[width=0.49\textwidth]{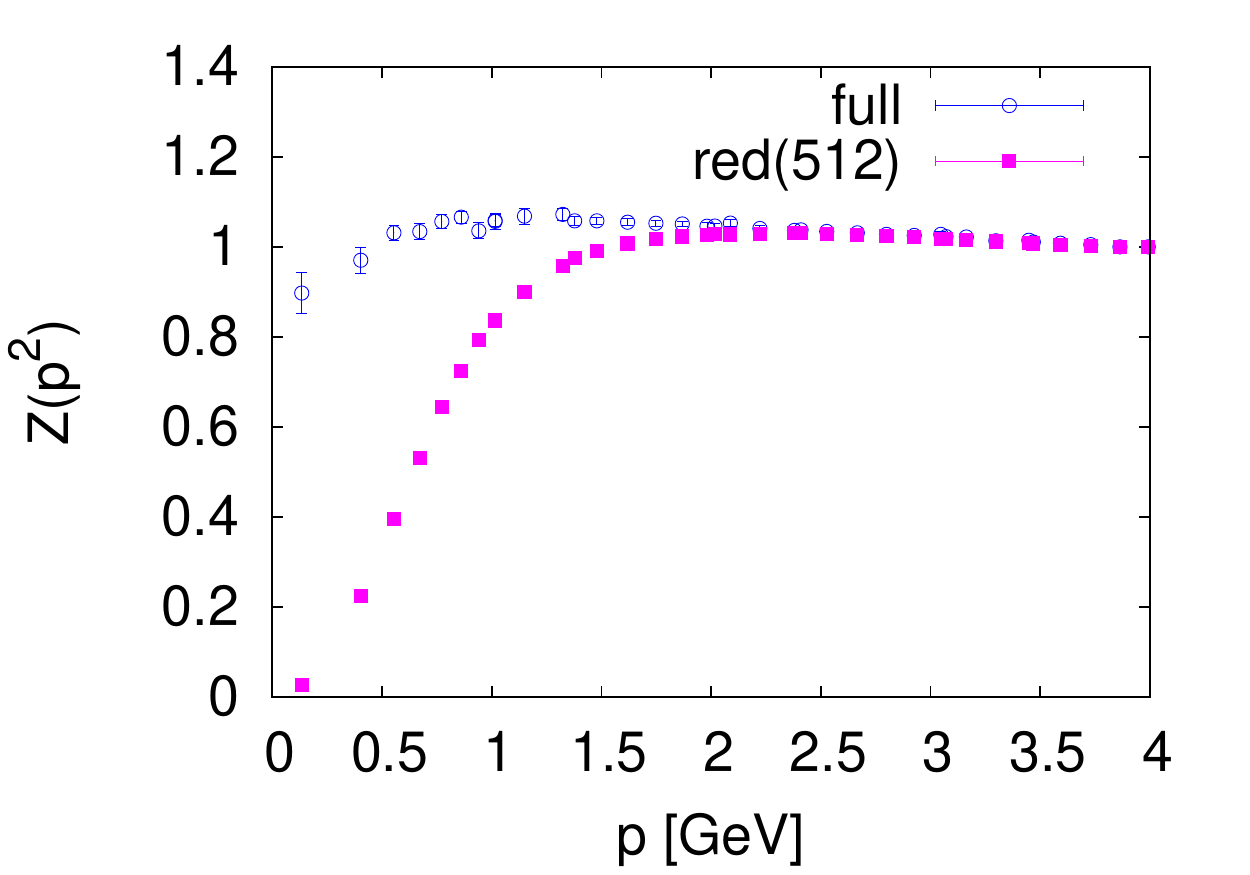}
	\caption{Left: The corrected mass function from the original theory 
		and after having subtracted the lowest 512 Dirac eigenmodes.
		Right: The same comparison for the wave-function renormalization
		function of the quark propagator.}
	\label{fig:MZ}
\end{figure}

We extract the quark wave-function renormalization 
function $Z_L(p)$ and the quark
mass function $M_L(p)$ from the reduced propagators of \eq{SRD} while varying the 
Dirac low-mode reduction level $k$ form 2 to 512.
We tree-level improve the modified propagators and apply the
aforementioned tree-level correction scheme.
The dressing functions from the full propagators together with the dressing functions
from the most extreme reduction (512 eigenmodes subtracted)
are presented in \fig{fig:MZ}.

The plot reveals amplification of infrared suppression of $Z_L(p)$
when subtracting Dirac low-modes while 
the range from medium to high momenta becomes not
affected at all.
Consequently,
far traveling quarks are suppressed in this framework.

\begin{figure}[htb]
	\center
	\includegraphics[width=0.75\textwidth]{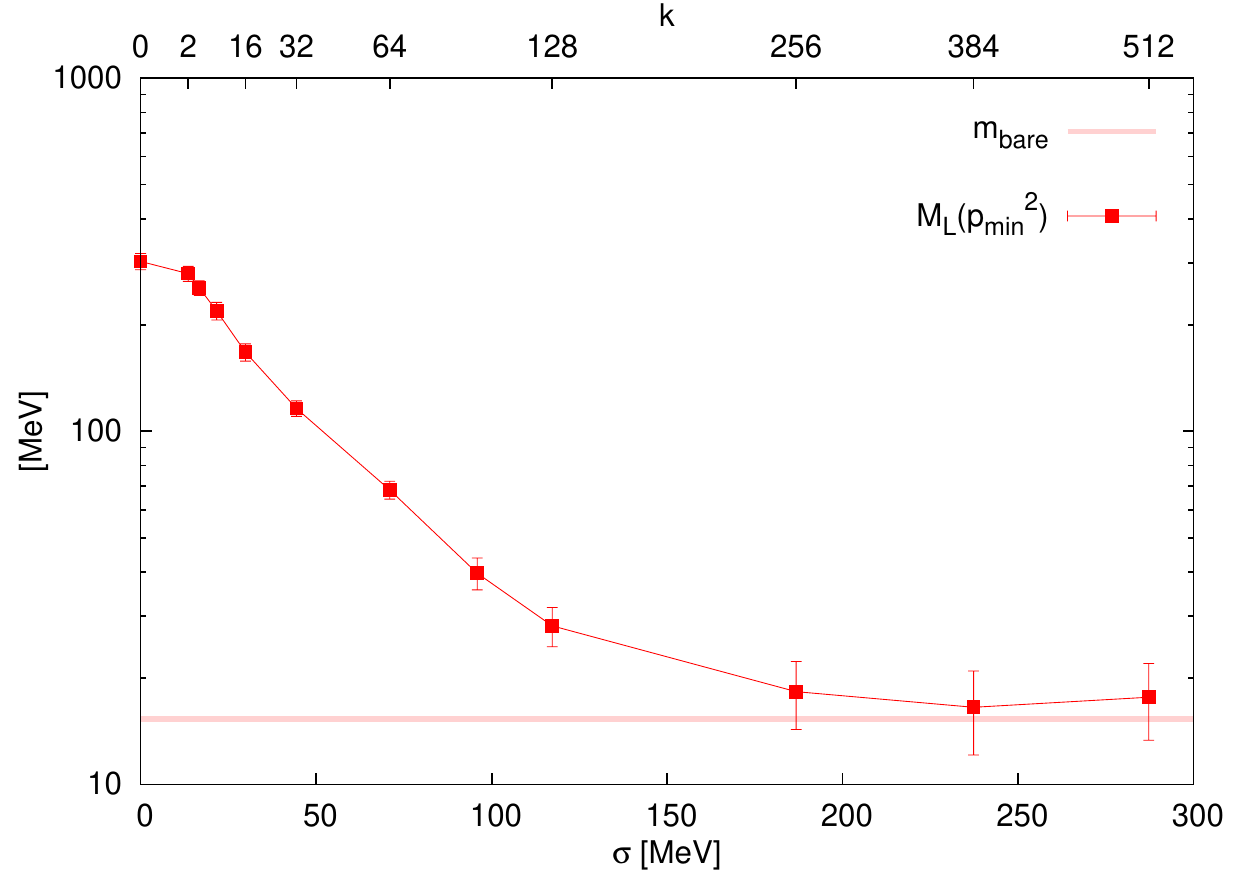}
	\caption{The value of the lattice mass function $M_L(p^2)$ at the lowest available momentum $p^2$
		as a function of the truncation level. The horizontal line is the value of the bare
		quark mass.}
	\label{fig:Mpmin}
\end{figure}

The mass function $M_L(p)$ gets strongly suppressed
in the infrared when removing the lowest eigenmodes, the dynamic
generation of mass ceases.
In \fig{fig:Mpmin} we plot the deep infrared mass of the CI quark propagator
from $M_L(p_{\textrm{min}}^2)$, at the smallest available momentum 
$p_{\textrm{min}}=\unit[0.13]{GeV}$, as a function of the Dirac low-mode reduction level.
We see that the value of the bare quark mass (horizontal line) is approximately reached 
at truncation level $k=256$.
Note that the reduction level $k$ can be translated to an energy scale which
is shown in the lower axis.

\section{Conclusions}
We found that the dynamically generated mass in the valence quark propagator vanishes
after a critical number of the lowest Dirac eigenmodes have been excluded from
the simulation. This is in accordance with the Banks--Casher relation which relates
the Dirac low-modes to the chiral condensate.
Thus, the rather large mass of hadrons which are built out of such truncated quark
propagators, see \fig{fig:mesons} and References~\cite{Lang:2011vw, Glozman:2012fj}
cannot be explained by the dynamical breaking of the chiral symmetry.
Interestingly, however, we find an intermediate Dirac low-mode truncation regime where
the dynamically broken chiral symmetry seems to be restored, as indicated by the degeneracy of
the masses of the chiral partner mesons $\rho$ and $a_1$ but, at the same time,
the quark propagator still shows a significant amount of dynamically generated mass.
Similar findings were obtained by the authors of \cite{OMalley:2011aa} 
who performed a hadron spectroscopy on gauge field configurations from which
center vortices have been cut out.

\begin{acknowledgments}
I thank C.~B.~Lang and L.~Ya.~Glozman for helpful discussions.
Support by the Research Executive 
Agency (REA) of the European Union under Grant Agreement 
PITN-GA-2009-238353 (ITN STRONGnet) is gratefully acknowledged.

\end{acknowledgments}


\providecommand{\href}[2]{#2}\begingroup\raggedright\endgroup

\end{document}